\def\Bbb#1{{#1\kern-.647em #1}}

\def \a{\alpha}
\def \b{\beta}
\def \g{\gamma}
\def \d{\delta}

\def \eps{\epsilon}
\def \om{\omega}

\def \be{\begin{equation}}
\def \eq{\end{equation}}

\def\tens{\mathop{\otimes}}
\def\id{\rm id}
\def\lform{\hbox{$\sqcup$}\llap{\hbox{$\sqcap$}}}
\def\eqn#1#2{\begin{equation}#2\label{#1}\end{equation}}
\def\proof{\goodbreak\noindent{\bf Proof\quad}}
\def\endproof{{\ $\lform$}\bigskip }
\def\o{{}_{(1)}}\def\t{{}_{(2)}}

\documentstyle[12pt]{article}

\newtheorem{prop}{Proposition}[section]
\newtheorem{lemma}[prop]{Lemma}
\newtheorem{thm}[prop]{Theorem}
\newtheorem{df}[prop]{Definition}

\newcommand \compl{C \! \! \! \! {\scriptscriptstyle
{}^{{}_|}}\ }

\def\Z{{\Bbb Z}}
\def\N{{\Bbb N}}

\def\A{A^{(n)}}
\def\P{P^{(n)}}
\def\AA{A^{(2n)}}
\def\PP{P^{(2n)}}

\def\h{{1/2}}
\def\n{{n/2}}
\def\kn{{kn/2}}
\begin{document}
\begin{titlepage}
\begin{center}
April 4, 1994
Revised: Jan 24, 1995
        \hfill  LBL-35452 \\
          \hfill    UCB-PTH-94/08 \\
\vskip .5in

{\large q -- deformed Dirac Monopole With Arbitrary
Charge} \vskip .5in
{Chong-Sun Chu, Pei-Ming Ho and Harold
Steinacker}\footnote
{email: hsteinac@physics.berkeley.edu}
\vskip .2in
{\em
Department of Physics \\University of California \\
and\\
   Theoretical Physics Group\\
    Lawrence Berkeley Laboratory\\
      University of California\\
    Berkeley, CA 94720}
\end{center}

\vskip .5in

\begin{abstract}
%{\bf Abstract}
 We construct the deformed Dirac monopole on the quantum
sphere for arbitrary charge using two different
 methods and show that it is a quantum principal bundle
in the sense of Brzezinski and Majid. We also give a
connection
 and calculate the analog of its Chern number by
integrating the curvature over $S^2_q$.
\end{abstract}

\end{titlepage}
\section{Introduction}

A major step towards a q -- deformed gauge theory is to
find a suitable concept of "quantum" fiber bundles.
Recently some versions of quantum bundles have been
proposed \cite{BM,Pf,Bud}, where both the base space and
the
fiber are quantum spaces. While all of them have their
particular advantages, only \cite{BM} also give a
detailed concept of a connection on a principal fiber
bundle. As examples, they
construct explicitely the q -- deformed Dirac Monopole
for charge 1 and 2, which are just $SU_q(2)$ resp.
$SO_q(3)$.
However, this construction is only formal for charge 1
and cannot be generalized to higher charges.

In this paper, we use the definition of \cite{BM} and
give 2 explicit constructions of the deformed Dirac
Monopole
$P^{(n)}$ for arbitrary "integer" charge $[n]$
with connection in the sense of \cite{BM}, both with
universal and a general calculus inherited from the 3-D
calculus
on $SU_q(2)$.
We first find $P^{(n)}$ as a certain subalgebra of
$SU_q(2)$, using a gradation on $SU_q(2)$.
In the second approach, $P^{(2n)}$ is constructed by
"glueing together", in the classical spirit, 2 trivial
bundles.
We give a condition when this can be done in general. It
is shown that the 2 constructions agree.
While the bundles are defined for any integer $n$, we
find local trivializations for even $n$ only;
for odd $n$, they would only be formal.
Thus we provide examples of nontrivial quantum principal
bundles which are not
quantum groups over some homogeneous space.
In the 3-D calculus, analogs of Chern numbers are
obtained by integrating the curvature two - form over
the
base $S^2_q$.

While the bundles are equipped with a star - structure,
the trivializations
respect this star - structure only in the classical
limit (as in \cite{BM}).
This appears to be a rather general feature of this
approach to quantum bundles.
While we take this as a feature rather than as a
problem, it seems that more examples and results are
needed
to find the most fruitful definitions.
Also, the connection one - forms are star - maps only in
the classical limit.
%We also point out that infinite series cannot be
avoided if one wants to consider general gauge
transformations.

\section{The Dirac Monopole as a Subalgebra of
$SU_q(2)$}

\subsection{Definition of quantum bundles}
Before giving the construction of the deformed monopole
bundle, let us summarize the definition and main
concepts
of quantum principal bundles in the sense of \cite{BM},
which we will use in this paper:

\begin{df}\label{principal.bundle}
\cite[Def 4.1]{BM}
 $P=P(B,A)$ is a quantum principal bundle (short: QPB)
with universal
differential calculus, structure quantum group $A$ and
base $B$ if
\begin{enumerate}
\item $A$ is a Hopf algebra
\item $(P, \Delta_{R})$ is a right $A$ -comodule
algebra; write $\Delta_R(p)=
 p^1\tens p^2 \in P\tens A$
\item $B=P^A = \{u \in P:\Delta_R u = u \otimes 1\}$
\item $(\cdot\tens\id)(\id\tens\Delta_R):P\tens P\to
P\tens A$ is a surjection
  (freeness condition)
\item $\ker\widetilde{\ }=\Gamma_{hor}$  (exactness
condition for the differential envelope)
\end{enumerate}
\end{df}
where horizontal forms $\Gamma_{hor}$ are defined by
\eqn{hor}{\Gamma_{hor} =  P(\Gamma_{B})P\subseteq
\Gamma_P}
and satisfy $\widetilde{\ } (\Gamma_{hor})=0$
identically.
The left $P$- module map $\widetilde{\  }$ is defined as
\eqn{tilde}{ \widetilde{\  }=(\cdot\otimes id) \circ
(id\otimes
\Delta_R)|_{P^2}:
\Gamma_P \rightarrow
P\otimes \ker \eps.}
In the dual picture, it generates the fundamental
(vertical) vector
fields on the bundle. We will use the same symbol
$\widetilde{\ }$ for the
extended map in condition 4. .

A connection on a quantum principal bundle is an
assignment of a left $P$-submodule
$\Gamma_{ver}\subseteq\Gamma_{P}$
such that:
\begin{enumerate}
\item $\Gamma_{P} = \Gamma_{hor} \oplus \Gamma_{ver}$,
\item  projection $\Pi : \Gamma_{P} \rightarrow
\Gamma_{ver}$ is right
invariant  i.e.
\begin{equation}
\Delta_{R} \Pi = ( \Pi \otimes id ) \Delta_{R},
\label{inv.conn}
\end{equation}
\end{enumerate}
A connection in $P$ is characterized by a right --
invariant left $P$ -- module map $\sigma:
P\otimes\ker\eps\rightarrow\Gamma_P$ splitting  the
exact sequence
\begin{equation}
0\rightarrow\Gamma_{hor}\rightarrow\Gamma_P\buildrel{\widetilde{\ }}
\over{\rightarrow}P\otimes\ker\eps\rightarrow 0,
\label{sequence}
\end{equation}
i.e. $\widetilde{\  }\circ\sigma = id$. The connection
form $\omega : A\rightarrow\Gamma_P$
is then given by
\begin{equation}
\omega(a) = \sigma (1\otimes (a-\eps(a))).
\label{def.omega.sigma}
\end{equation}
Conversely,  $\sigma(p\tens a)=p\omega(a)$ for $p\tens
a\in P\tens \ker\eps$.

In order to use a general calculus, the above
definitions have to be augmented.
As usual, the first -- order calculus
on $A$ shall be determined by a right ideal
$M_A\subset\ker\eps$ as $\Gamma_A =A^2/N_A$, where $N_A
= \kappa(A\otimes M_A)$
and the map $\kappa : A\tens A \rightarrow A\tens A$ is
given by
\eqn{kappa.A}{\kappa (a\otimes a') =
\sum aSa'\o\otimes a'\t ,}
where the Sweedler's notation \cite{S} is employed.

Similarly on $P$, one assumes that the first -- order
differential structure $\Gamma_P$ is a quotient
of the universal one, $\Gamma_P=P^2/N_P$ where $N_P$ is
a sub-bimodule of $P^2$. The calculus for higher orders
is then
uniquely determined. In order to have consistent calculi
on $P$ and $A$, the
definition of a QPB is:

\begin{df}\label{gen.principal.bundle}
\cite[Def 4.9]{BM}
 $P=P(B,A,N_P,M_A)$ is a quantum principal bundle with
structure quantum group
 $A$, base $B$ and quantum differential calculi defined
by $N_P$, $M_A$ if
\begin{enumerate}
\item $A$ is a Hopf algebra
\item $(P, \Delta_{R})$ is a right $A$ - comodule
algebra
\item $B=P^A = \{u \in P:\Delta_R u = u \otimes 1\}$
\item $(\cdot\tens\id)(\id\tens\Delta_R):P\tens P\to
P\tens A$ is a surjection
  (freeness condition)
\item $\Delta_R N_P \subset N_P\otimes A$ (right
covariance of differential
structure).
\item $\widetilde{\  }(N_P)\subset P\otimes M_A$
(fundamental vector fields
compatibility condition)
\item $\ker\widetilde{\  }_{N_P}=\Gamma_{hor}$
(exactness condition).
\end{enumerate}
\end{df}
where $\widetilde{\ }_{N_P}$ is the map induced by
$\widetilde{\ }$.
A connection on a QPB with general calculus is again
determined by a splitting
$\sigma$ of the sequence
\begin{equation}
0\rightarrow\Gamma_{hor}\rightarrow\Gamma_P\buildrel{\widetilde{\ }_{N_P}}
\over{\rightarrow}P\otimes\ker\eps/_{M_A}\rightarrow 0 .
\label{gen.sequence}
\end{equation}
Point (6) may be replaced by the slightly stronger
condition \cite{BM}

6.' $\quad \widetilde{\  }(N_P) = P\otimes M_A$,
which we will adopt in section (3)ff.

\subsection{Dirac monopole with universal calculus}

$SU_q(2)$ is generated as usual by $\a,\b,\g,\d$ with
the commutation relations
\[\alpha\beta = q\beta \alpha , \quad \alpha\gamma = q
\gamma\alpha . \quad \alpha \delta = \delta\alpha +
(q-q^{-1})\beta\gamma, \]
\[\beta\gamma = \gamma\beta, \quad \beta\delta = q\delta
\beta
,\quad \gamma \delta = q \delta \gamma \]
and a determinant relation
$\alpha\delta-q\beta\gamma=1$. The $*$-structure is
$\alpha^*=\delta$, $\beta^*=-q\gamma$.

It is shown in \cite{BM} that $P^{(1)}=SU_q(2)$ and
$P^{(2)}=SO_q(3)$ are QPB 's with structure quantum
group
$A=k<Z^{1/2}, Z^{-1/2}> = U(1)$ resp. $A=k<Z, Z^{-1}> =
U(1)$ ($k = \compl$ in order to recover the classical
monopole for $q = 1$), right coaction
$\Delta_R: P \rightarrow P\tens A$ defined by

\be \Delta_R\pmatrix{\alpha&\beta\cr
\gamma&\delta}=\pmatrix{\alpha\tens Z^\h&\beta\tens
Z^{-\h}\cr
\gamma\tens Z^\h &\delta\tens Z^{-\h}}
\eq
and base $B=SU_q(2)^A = <1,b_-=\a\b,b_+=\g\d,b_3=\a\d
>$. $SO_q(3)$ is defined here
as the even elements of $SU_q(2)$.

To describe Dirac monopoles with higher charges, we
define a degree for a monomial in $SU_q(2)$ as follows:
\be
 deg(\a^a \b^b \g^c \d^d) =a+c-b-d
\eq
irrespective of ordering. We shall show that the
q-deformed Dirac monopole with (classical) charge
$n\in\N$ is
\be
\P=<\{ p \in SU_q(2), deg (p) =nk, k \in \Z \}>.
\label{Pn}
\eq
and
$\A=k<Z^{n/2}, Z^{-n/2}>$.
The superscript denotes the charge of the bundle. So
monomials in $\P$ have a degree which is
an arbitrary multiple of $n$.
For $n=2$ we have $P^{(2)} = SO_{q}(3)$, in agreement
with \cite{BM}.
The coaction $\Delta_R:\P\rightarrow \P \tens \A$ is
induced from the above as a star - algebra map and is
compatible with the grading, and $B=P^{(n)A}=<\{p \in
\P, deg(p)=0\}>=S^2_q$.
Also note that the above star - structure carries over
to $\P$.

We do not have to introduce trivializations and "local"
bundles here, this will be done in the second
approach. However to motivate the above definitions, let
us remark that e.g. $\Phi_0(Z^{\pm
n/2})=(\d^{-1}\a)^{\pm n/2}$
will turn out to be a local trivialization (at least for
$q=1$); since locally $P=B\tens A$, this gives the
characterization above.
In the classical limit, $\A$ is nothing but the
functions on $U(1)$ with $n$-th roots of unity
identified,
so the winding number of $\P$ will just be $n$.

For $n \geq 2$, we can now show the following:

\begin{prop}
$\P(B, \A)$ is a QPB with universal calculus.
\end{prop}
\proof

For condition 5. of definition 2.1, we see that $\ker
\widetilde{\  }=\Gamma_{hor}^{(1)} \cap \Gamma_P^{(n)} =
\Gamma_{hor}^{(n)}$,
since $deg(B)=0$.

To show 4., consider the
monomial $p \tens Z^{nk/2} \in \P \tens \A \subset
P^{(1)} \tens A^{(1)}$ for $k \in \Z$.
Since $ P^{(1)}$ is a QPB, there exists $\sum p_{i_1}
\tens p_{i_2}
\in P^{(1)} \tens P^{(1)}$ such that
$\widetilde{\  } (\sum p_{i_1} \tens p_{i_2})= p\tens
Z^{nk/2}$. Now,
$\widetilde{\  } (\sum p_{i_1} \tens p_{i_2})=
\sum p_{i_1} p_{i_2} \tens Z^{deg(p_{i_2})/2}$.
Therefore, $deg(p_{i_2})=nk$, and $\space p_{i_2} \in
\P$  for all $i_2$.
Also, $deg(p_{i_1} p_{i_2})=deg(p)$ and $p \in \P,$ so
$deg(p_{i_1}) \in n\Z$.
Hence, $p_{i_1} \in \P$  for all $i_1$. Surjectivity is
proved.

\endproof

A possible connection one -- form on $\P$ is given by
\begin{eqnarray}
\omega(Z^{kn/2})=S((\a^{kn})_{(1)})d (\a^{kn})_{(2)} =
\kappa(1\tens (\a^{kn}-1)), \\
\omega(Z^{-kn/2})=S((\d^{kn})_{(1)})d (\d^{kn})_{(2)} =
\kappa(1\tens (\d^{kn}-1))
\end{eqnarray}
for $k>0$, where $\kappa$ is defined in \cite{BM}.
This $\omega$ is well - defined, since
$S((\a^{kn})_{(1)})$, $S((\d^{kn})_{(1)})$,
$(\a^{kn})_{(2)}$, $(\d^{kn})_{(2)} \in P^{(n)}$.
This connection was found observing that this is the
trivial connection \cite{BM}
obtained form the trivialization $\Phi(Z^{n/2}) = \a^n$,
which is a gauge - transformation of the trivialization
(\ref{triv_0}) used in section 4.
Note that the above trivialization does not respect the
star - structure even for $q=1$,
nevertheless it is useful to e.g. find a connection; in
the 3D - calculus, it simplifies to
(\ref{connect}), and for even $n$, we would have
obtained the same
$\omega$ using  (\ref{triv_0}). Quite generally, gauge
- transformations tend to
spoil the star - structure (and algebra - structure, as
pointed out in \cite{BM})
of a trivialization.

To prove that $\omega$ defines a connection, we use
Proposition 4.4 in \cite{BM}. We have to show
\begin{enumerate}
\item $\omega(1)=0$
\item $\widetilde{\  }\omega(a)=1\tens a-1\tens
1\eps(a)$ for all $a\in A$
\item $\Delta_R\circ\omega=(\omega\tens\id)\circ Ad_R$
\end{enumerate}
1. is obvious, 2. holds since for $k>0$,
\begin{eqnarray}
\widetilde{\ } \om(Z^\kn)
&=& S((\a^\kn)_{(1)})  (\a^\kn)_{(2)}^{(1)} \tens
(\a^\kn)_{(2)}^{(2)} -1\tens 1
\nonumber \\
&=& S((\a^\kn)_{(1)})  (\a^\kn)_{(2)} \tens Z^\kn
-1\tens 1 \nonumber\\
&=& 1 \tens  (Z^\kn -1)
\end{eqnarray}
and similarly for $k<0$ as is easily seen from our
coaction. For 3.,
\begin{eqnarray}
\Delta_R \om (Z^\kn) &=& S((\a^\kn)_{(1)}) d
(\a^\kn)_{(2)}\tens Z^{-kn/2}Z^{kn/2} \nonumber \\
&=& \om(Z^\kn) \tens 1 = (\om\tens id) Ad_R(Z^\kn)
\end{eqnarray}
 and similarly for $k<0$.

\subsection{General Calculus}

It is also shown in \cite{BM} that
$P^{(1)}(B,A^{(1)}, N_P^{(1)}, M_A^{(1)})$ is a QPB with
general calculus
where $N_P^{(1)}$ defines the 3-D calculus on $SU_q(2)$,
i.e. the right
ideal $M_P^{(1)}$ is generated by the six elements
\begin{equation}
\d+q^{2}\a-(1+q^{2}), \g^{2}, \b\g, \b^{2}, (\a-1)\g,
(\a-1)\b
\end{equation}
and
\be M_A^{(1)}= \pi (M_P^{(1)})
=<\{Z^{-1/2}+q^{2}Z^{1/2}-(1+q^{2})\}>. \eq

The projection $\pi:P\rightarrow A$ (dual of $U(1)
\subset SU(2)$ )
is an algebra map
\begin{equation}
\pi\left(\begin{array}{ll}
     \a&\b\\
     \g&\d
    \end{array}\right)=\left(\begin{array}{ll}
                   Z^{1/2}&0\\
                   0&Z^{-1/2}
                  \end{array}\right).
\end{equation}

For the general case, take $N_P^{(n)} =N_P^{(1)} \cap
(\P)^2$, $M_A^{(n)}$ generated by
$ Z^{-n/2} +q^{2n} Z^\n -(1+q^{2n})$, i.e. $Z^{n/2}d
Z^{n/2}=q^{2n} d Z^{n/2} Z^{n/2}$
or equivalently $N_A^{(n)} =N_A^{(1)} \cap (\A)^2$. Then
we have:

\begin{prop}
$P^{(n)}(B,A^{(n)}, N_P^{(n)}, M_A^{(n)})$ is a QPB.
\end{prop}
\proof
Conditions 1-4 of definition 2.2 are trivial as $\P(B,
\A)$ is a QPB. Condition 5 is obvious
because of our simple $\Delta_R$. Now, notice that
$P^{(1)}(B,A^{(1)}, N_P^{(1)}, M_A^{(1)})$ is a QPB, so
$\ker \widetilde{\  }_{N_P^{(n)}}
=(\ker \widetilde{\  }_{N_P^{(1)}}) \cap \Gamma_P^{(n)}
= \Gamma_{hor}^{(1)} \cap \Gamma ^{(n)}_P
%= (\Gamma_{hor}^{(1)} \cap \Gamma_P^{(n)}) / (N^{(1)}_P
\cap \Gamma ^{(n)}_P)
= \Gamma_{hor}^{(n)}$ since $deg($B$)=0$.
For the same reason, we know that
$\widetilde{\  }N_P^{(1)} \subset P^{(1)} \tens
M_A^{(1)}$. Therefore,
$\widetilde{\  }N_P^{(n)} \subset P^{(1)} \tens
M_A^{(1)}$.
But $\widetilde{\  }(p_1dp_2)=p_1p_2^{(1)} \tens
p_2^{(2)}- p_1 p_2 \tens 1
\in \P \tens \A$, for all $p_1, p_2 \in \P,$
so $\widetilde{\  }N_P^{(n)} \subset P^{(n)} \tens
M_A^{(n)}$.
Hence we see that
$P^{(n)}(B,A^{(n)}, N_P^{(n)}, M_A^{(n)})$ is a QPB.
Also, note that the 3D - calculus respects the star -
structure \cite{W3}.
\endproof

The above connection one -- form $\omega$ defines also a
connection on $\P$ with our general
calculus. The only thing that remains to be checked
according to Proposition 4.10 of \cite{BM} is
$\omega(M_A^{(n)})=0$. But this is clear since $\om
(Z^{-n/2}+q^{2n} Z^\n -(1+q^{2n}) )=
\kappa(1\tens (\d^n+q^{2n} \a^n-(1+q^{2n})))
\in\kappa(1\tens M_P^{(n)})$.

Thus, $\om$ is a connection form on our bundle and is
given by
\begin{eqnarray}
\om(Z^{k\n})  &=& [kn]_{q^{-2}} \om ^1, \nonumber \\
\om(Z^{-k\n}) &=&-[-kn]_{q^{-2}} \om ^1 =
-q^{2kn}\om(Z^{k\n}) \label{connect}
\end{eqnarray}
if viewed in $SU_q(2)$, where $[n]_q =
\frac{q^n-1}{q-1}$.
This generalizes the result of \cite{BM} for
$n=1$ and 2. Since $(\omega^1)^{\ast} = -\omega^1$,
$\omega$ is a star - map for $q=1$ only. We have used
$\om^1 \a=q^{-2} \a \om^1, \om^1 \g=q^{-2} \g \om^1$,
where
$\om^1 =\d d \a -q^{-1} \b d \g$ is a left invariant
form in $SU_q(2)$.

To our knowledge, (\ref{Pn}) is also a new description
of the classical Dirac monopole.

\section{Combining patches to bundles}

Let us now show how nontrivial QPB's can be obtained by
"glueing" together "local" bundles.
To avoid repeating ourselves too much, we will give the
following statements for the case of a general calculus
only; the universal calculus
is recovered by putting $M_A=N_A=N_P=0$.
We first observe that the conditions 4. and 7. in
definition \ref{gen.principal.bundle}
are equivalent to the exactness of the
sequence (\ref{gen.sequence}).
%
%\begin{equation}
%0\rightarrow\Gamma_{hor}\rightarrow\Gamma_P\buildrel{\widetilde{\ }_{N_P}}
%\over{\rightarrow}P\otimes\ker\eps/_{M_A}\rightarrow 0
%\label{gen.sequence}
%\end{equation}
%
%where $j$ is the inclusion of $\Gamma_{hor}$ into
$\Gamma_P$.

\begin{lemma}\label{gen.bundle}
$P(B,A,N_P,M_A)$ satisfying conditions 1. to 3.,5. and
6.' of the
definition \ref{gen.principal.bundle} is a QPB with
general calculus if
and only if the sequence (\ref{gen.sequence}) is exact.
\end{lemma}
\proof
Exactness of (\ref{gen.sequence}) at $\Gamma_P$ is just
the condition 7. above.

Assume first $P$ is a QPB. Then by condition 4. , for
any $p\tens a \in P\tens \ker\eps/M_{A}$ there exists
$p_1 \tens p_2 \in P\tens P$ with $\widetilde{\
}(p_1\tens p_2)=p_1p_2^1\tens p_2^2 = p\tens a$.
Applying $\id\tens\eps$ to this equation we get
$0=p_1p_2^1\eps(p_2^2) = p_1 p_2$, i.e. $p_1\tens p_2
\in \Gamma_P$, which shows that $\widetilde{\ }$ in
(\ref{gen.sequence}) is surjective, so it is exact.

Conversely, suppose (\ref{gen.sequence}) is exact. Take
any $p\tens a =
p\tens(a-\eps(a)) + p\tens \eps(a) \in P\tens A$ .
Since $\widetilde{\ }_{N_P}$ is surjective, there exists
$p_1 dp_2 \in\Gamma_P$
with $\widetilde{\ }_{N_P}(p_1 dp_2)=p\tens(a-\eps(a))+
P\tens M_A$.
Now $\widetilde{\ }(p\tens \eps(a))=p\tens\eps(a)$ and
from 6'.
$\widetilde{\ }(N_P)=P\tens M_A$, so condition 4. is
satisfied.
\endproof

Assume now we have 3 quantum principal bundles
$P_0(B_0,A,N_0,M_A), \newline
P_1(B_1,A,N_1,M_A) \subset
P_{01}(B_{01},A,N_{01},M_A)$ ($P_{01}$ corresponds to
the bundle on the "overlap" $B_{01}$
of $B_0$ and $B_1$) and we would
like to know if $P_0$ and $P_1$ can be understood as two
patches of a "global" quantum bundle $P(B,A,N,M_A)
\subset P_0,P_1 \subset P_{01}$.
A natural guess is that $P=P_{0}\cap P_{1}$.
In this case the coactions $\Delta_{R_i} :
P_i\rightarrow P_i\tens A$ certainly
must agree in $P$. If we want a connection on $P$, then
we should also have
connection forms $\omega_i: A \rightarrow \Gamma_{P_i}$
which agree on the overlap, i.e.
$\omega_0(a)=\omega_1(a)$
in $\Gamma_{P_{01}}$.

However, some care must be taken if we want to compare
differential forms on different
patches. First of all, the differential structures on
$P_i$ must be compatible, i.e. we should have
$N_0=N_{01}\cap P_0^2$, $N_1=N_{01}\cap P_1^2$ and
$N_P\equiv N=N_{01}\cap P^2=N_0\cap N_1$. But this is
not enough:
Suppose we have any 2 differential forms -- not
necessarily connections --  $\omega_0 \in \Gamma_{P_0}$
and $\omega_1 \in \Gamma_{P_1}$ and find by doing
calculations in $\Gamma_{P_{01}}$ that they are equal.
One would certainly like to conclude, as in the
classical case, that they determine a "global" form
$\omega$ in $\Gamma_P$.
This is not evident, it is a condition on the calculus.
It motivates the
following definition:
The above calculi on $P_0, P_1, P_{01}$ are called {\em
admissible} if
\be
\omega_0 = \omega_1 +n_{01}\quad {\rm for} \quad
\omega_i \in \Gamma_{P_i} \label{ad1}
\eq
implies that there exists a $\omega\in\Gamma_P$ such
that
\be
\omega=\omega_0+n_0 = \omega_1 + n_1 , \quad n_i \in
N_i. \label{ad2}
\eq
In other words, $\omega_0=\omega_1$ determines a $\omega
\in
\Gamma_{P_0}\cap\Gamma_{P_1} = \Gamma_{P}$, where the
intersection is defined
as intersection of the cosets.

A calculus which does not satisfy this condition would
be highly unpracticable for global statements.
The universal calculus is certainly admissible since
$(P_0\tens P_0) \cap (P_1\tens P_1) = P\tens P$ implies
$\Gamma_{P_0} \cap \Gamma_{P_1} = \Gamma_P$.
The calculus we will consider on the monopole -- bundle
will be shown to be admissible too, using a fairly
general line of reasoning.

\begin{thm} \label{gen.patch}
In the above situation, $P=P_0 \cap P_1=P(B,A,N,M_A)$ is
a quantum principal bundle with base
$B=B_0\cap B_1$ and connection
if we have admissible differential structures which
satisfy $\widetilde{\ }(N)
=P\otimes M_{A}$, connection forms $\omega_0 = \omega_1$
on $P_0$ resp. $P_1$,
and $\Gamma_{0 hor} \cap \Gamma_{1 hor} = \Gamma_{hor}.$
Conversely, if $P=P_0 \cap P_1$ is a quantum principal
bundle,
then $\Gamma_{0 hor} \cap \Gamma_{1 hor} =
\Gamma_{hor}.$
\end{thm}

\proof
First, $\Delta_R(p)\in P_0\tens A\cap P_1\tens A =
P\tens A$ for $p\in P$ implies 2. in Def
\ref{principal.bundle}.
Further, $B=P^A=(P_0\cap P_1)^A=P_0^A\cap P_1^A=B_0\cap
B_1$.
By the above definition of the differential structures
condition 5. is satisfied,
since $\Delta_{R_i} : P_i\rightarrow P_i\tens A$ do not
"leave" the bundles.

Assume $\Gamma_{0 hor} \cap \Gamma_{1 hor} =
\Gamma_{hor}$.
Since $\omega_0(a)=\omega_1(a)$ and the calculus is
admissible, this defines
$\omega(a)\in\Gamma_P$ and $\sigma(p\tens a)=
p\omega(a) \in \Gamma_P$ for $(p\tens a) \in
P\tens\ker\eps$.
{}From proposition 4.10 in \cite{BM} it follows that
$\omega$ is a connection 1 -- form.
Now $\widetilde{\ } \sigma(p\tens a)=p\tens a$
shows that the map $\widetilde{\ }$ in
(\ref{gen.sequence}) is surjective.

It remains to show $\ker\widetilde{\ }=\Gamma_{hor}$.
Let $p_1d p_2
\in\Gamma_P$. Since $P_0$ and $P_1$ are quantum bundles,
$\widetilde{\ }(p_1 dp_2)=0$ implies $p_1 dp_2 \in
\Gamma_{0 hor}\cap\Gamma_{1 hor} = \Gamma_{hor}$ by
assumption.
Now Lemma ~\ref{gen.bundle} tells us that $P(B,A,N,M_A)$
is a quantum principal bundle.

Conversely, assume $P=P_0\cap P_1$ is a quantum
principal bundle. Let
$p_1 dp_2 \in \Gamma_{0 hor}\cap\Gamma_{1 hor}$.
Then $\widetilde{\ }(p_1 dp_2) =0$. Since $\Gamma_{0
hor}\cap\Gamma_{1 hor} \subset
\Gamma_{P_0}\cap\Gamma_{P_1}
=\Gamma_P$ and $P$ is a QPB, this implies
$p_1 dp_2 \in \Gamma_{hor}$. The other inclusion
$\Gamma_{0 hor}\cap\Gamma_{1 hor} \supset \Gamma_{hor}$
is trivial.
\endproof

%"Admissible" is not needed if $\omega(a) \in \Gamma_P$
explicitely.
If there are several "patches" $P_i$, then the above
theorem generalizes
inductively in an obvious way.
One can show that if $\widetilde{\ }(N_P)=P\tens M_A$,
then
$\Gamma_{0 hor}\cap \Gamma_{1 hor} =\Gamma_{hor}$
follows from $\Gamma^u_{0 hor}\cap \Gamma^u_{1 hor}
=\Gamma^u_{hor}$ (universal calculus). More generally,
we have

\begin{lemma} \label{univ.gen}
If $P(B,A)$ is a QPB with universal calculus and we have
$N_P$ and $M_A$
satisfying conditions 5. and 6'. of definition
\ref{gen.principal.bundle},
then $P(B,A,N_P,M_A)$ is a QPB with general calculus.
Conversely, if
$P(B,A,N_P,M_A)$ is a QPB and $\widetilde{\ }(n)=0$ for
$n\in N_P$ implies
$n\in \Gamma^u_{hor}$, then $P(B,A)$ is a QPB with
universal calculus.
\end{lemma}

\proof
First suppose $P(B,A)$ is a QPB; we have to show that
$\ker\widetilde{\ }_{N_P}
\subset \Gamma_{hor}$. Let $\widetilde{\ }_{N_P}(\g)=0$.
This means
$\widetilde{\ }(\g)\in P\tens M_A = \widetilde{\ }(N_P)$
by 6'. . So there is
a $n \in N_P$ with $\widetilde{\ }(n-\g)=0$. But $P$ is
a QPB with universal
calculus, so it follows $\g\in\Gamma^u_{hor} +n$, i.e.
$\g\in\Gamma_{hor}$.

The converse statement can be proved similarly.
\endproof

\section {Dirac Monopoles by Patching Two Trivial QPBs}

We can now present the second construction of the Dirac
monopoles for general calculus
as an illustration of the general method above.
This will be done for even "charge" only; for odd
charge, the trivializations etc.
would only be formal.

We define two trivial QPBs $\PP_{0}$ and $\PP_{1}$, and
then show that
$\PP=\PP_{0}\cap \PP_{1}$ is the monopole of charge
$2n$.

For $\PP_0$, as motivated by the charge 2 case in
\cite{BM}, we now try to define
the base $B_{0}$, fiber $\AA$ and trivial bundle
$\PP_{0}$ be specified by their generators as:
\begin{eqnarray} \label{B0}
B_{0}&=&<\{1,b_{-},b_{+},b_{3},(b_{3}+q^{2m}-1)^{-1};\quad m\in\Z \}>, \\
   \AA&=&<\{Z^n,Z^{-n}\}>,\\
\PP_{0}&=&<B_{0}\cup \{(\d^{-1}\a)^n,(\a^{-1}\d)^n\}>.
\end{eqnarray}
We are going to show that they give a trivial QPB.

The commutation relations between the generators of
$P_{0}$ are induced by
$SU_{q}(2)$ through the following expressions
\cite{Podles}:
\begin{eqnarray}
b_{-}=\a\b,&
b_{+}=\g\d,&
b_{3}=\a\d \label{base-SU2},
\end{eqnarray}
where $\a,\b,\g,\d$ are generators of $SU_{q}(2)$ with
the
well-known relations stated before. The commutation
relations involving inverses are
obtained by multiplying them from both sides by inverses
of generators.

In the classical limit $q = 1$, $B_{0}$ becomes the
algebra of the
functions on $S^{2}\backslash$\{south pole\}, and
$b_{\pm}=\pm(x\pm iy)$,
$b_{3}=z+1/2$, where $x,y,z$ are the Cartesian
coordinates. Note that
$\a\d-\b\g=1$ is equivalent to
$x^{2}+y^{2}+z^{2}=(1/2)^{2}$. The somewhat complicated
definition here (see \cite{BM_err}) will become clear
below.
$\PP_0$ as a trivial bundle is generated by the base
$B_0$ and the fibers, cp.
(\ref{triv_0}).

Define a coaction $\triangle_{R}$ on $\PP_{0}$ such that
$B_{0}=(\PP_{0})^{\AA}$:
\begin{eqnarray}
\triangle_{R}(1)&=& 1\otimes 1, \\
\triangle_{R}(b_{i})&=&b_{i}\otimes 1, \quad i=-,+,3, \\
\triangle_{R}((\d^{-1}\a)^{\pm n})&=&(\d^{-1}\a)^{\pm
n}\otimes Z^{\pm n}.
\end{eqnarray}

The trivialization $\Phi_{0}$ is defined as
\be
\Phi_{0}(1)=1, \qquad
\Phi_{0}(Z^{\pm n})=(\d^{-1}\a)^{\pm n}  \label{triv_0}
\eq
which generalizes the trivialization in \cite{BM}.
To see that we have a trivial QPB, we first have to show
that $B_0$ is the
invariant subalgebra of $\PP_0$ under the above
coaction. This is clear if any
$p_0 \in \PP_0$ can be written as a sum of terms $B_0
(\d^{-1}\a)^{kn}$. Thus we must be able to commute
$B_0$ through $(\d^{-1}\a)$. Writing down the
commutation relations explicitely, one can always obtain
relations like
$\a B_0 \a^{-1} \in B_0$. Note that for $m \in \Z$,
$q^{-2m}\a^{-m}b_3^{-1}\a^m =
q^{-2m}\d^{m}b_3^{-1}\d^{-m} =
(b_{3}+q^{2m}-1)^{-1}$ (cp. \cite{BM_err}) and  so in
general,
\be
q^{-4nk} (\a^{-1} \d)^{nk}(b_3+q^{2m}-1)^{-1} (\d^{-1}
\a)^{nk} =  (b_3+q^{4nk+2m}-1)^{-1},
\quad k \in \Z.
\eq
This shows that $B_0$, as defined in (\ref{B0}) is the
invariant subalgebra,
and one can also see the necessity to include all the
generators of $B_0$.

$\Phi_{0}$ is convolution-invertible with
$\Phi_{0}^{-1}(Z^{\pm n})
=(\d^{-1}\a)^{\mp n}$, and is also an intertwiner:
$\triangle_{R}\circ\Phi_{0}=(\Phi_{0}\otimes
id)\circ\triangle_{A}$, where
$\triangle_{A}(Z^n)=Z^n\otimes Z^n \space$ is the
coproduct on $\AA$.
So $\PP_{0}$ is a trivial QPB.

Below we will need the following alternative
representation of $\PP_0$:
\be
\PP_0=\{p\in <SU_q(2) \cup \{(\d\a)^{-1},(\a\d)^{-1}\}>:
deg(p)=2kn, k \in \Z\}\equiv \tilde\PP_0
\label{rep_P0},
\eq
i.e. the algebra generated by $SU_q(2)$ and
$(\d\a)^{-1}, (\a\d)^{-1}$, with degrees
being multiples of $2n$. To see this, note that
$\PP_0 \subset \tilde\PP_0$ because $\d^{-1}\a =
(\a\d)^{-1}\a^2$ etc. and
$b_3^{-1} = (\a\d)^{-1}$, so $\a^{-n} b_3^{-1}\a^n \in
\tilde\PP_0$ also.
To see the other inclusion, we first show that $B_0$ is
also the invariant
subalgebra (under the coaction of $\AA$)  of
$\tilde\PP_0$: we have just seen
$B_0 \subset \tilde\PP_0$, and the same commutation
relations as above show that indeed
$B_0 = (\tilde\PP_0)^{\AA}$. But this means that
$\tilde\PP_0$ is a QPB with the same trivialization
$\Phi_0$ as above. Thus we
know (from \cite{BM} Example 4.2) that $\tilde\PP_0 =
B_0 \Phi_0(\AA) = \PP_0$.

The discussion on $\PP_1$ is parallel to that on
$\PP_0$, but much easier.
Therefore we just give the relevant equations:
\begin{eqnarray}
B_{1}    &=&<\{1,b_{-},b_{+},b_{3},(b_{3}-1)^{-1}\}>, \\
\PP_{1}&=&<B_{1}\cup \{(\g\b^{-1})^n,(\b\g^{-1})^n \}>,
\\
\AA      &=&<\{Z^{n},Z^{-n}\}>,\\
\triangle_{R}(b_{i})&=&b_{i}\otimes 1, \quad i=-,+,3, \\
\triangle_{R}((\g\b^{-1})^{\pm n})&=&(\g\b^{-1})^{\pm
n}\otimes Z^{\pm n}, \\
\Phi_{1}(Z^{\pm n})&=&(-\g\b^{-1})^{\pm n}
\end{eqnarray}
and $\PP_1$ is also a trivial QPB.
Note again that $deg(B_1)=0$ and
$deg(\Phi_i(Z^{n}))=2n.$

The "overlap" $\PP_{01}$ of $\PP_0$ and $\PP_1$ is
similarly defined by
\begin{eqnarray}
B_{01}    &=&<B_0 \cup \{(b_{3}-1)^{-1}\}>, \\
\PP_{01}&=&<B_{01}\cup \{(\g\b^{-1})^{\pm
n},(\d^{-1}\a)^{\pm n}\}>
\end{eqnarray}
and so on as above. On $\PP_{01}$, both trivializations
can be used, with the transition function
\be
\g_{01}(Z^{n})=\Phi_0(Z^{n})\Phi_1^{-1}(Z^{n})=
                 (-q^2 b_3^{-1}b_-^2(b_3-1)^{-1})^n \in
B_{01} .  \label{gt}
\eq
It should be noted that while these trivial bundles are
closed under the star -
operation, the maps $\Phi_i$ respect this star -
structure only for $q=1$.
This appears to be very hard to avoid in this framework,
and we accept it here.

Now define the Dirac - monopole bundle with charge $2n$
by
\be
 \PP=\PP_{0}\cap \PP_{1}.
\eq
We will now show that for even charges this construction
agrees with the one in section 2.
First, we prove
\begin{prop} \label{intersect}
\be
\PP =<{p\in SU_{q}(2): deg(p)=2nk,k\in \Z}>.
\eq
\end{prop}

\proof
Let $p_0$, $p_1 \in \PP_0$ resp. $\PP_1$ and $p_0=p_1$.
Note that $\b$, $\b^{-1}$, $\g$, $\g^{-1}$ can be
commuted through any terms by just picking up powers of
$q$.
Multiplying $\a^{-1}$,$\d^{-1}$ to the relation
$\a\d=\d\a+(q-1)(\frac{q+1}q)\b\g$ appropriately from
both sides, one gets relations
like $\d\a^{-1}=\a^{-1}\d + (q-1)(...)$ and
$\a^{-1}\d^{-1} =\d^{-1}\a^{-1} + (q-1)(...)$, i.e.
{\em one can order thing in any way up to terms
proportional to $(q-1)$}.

Let us define a {\em normal form for $p_1$} as follows:
bring all $\b, \g$ to the right of all $\a, \d$
and order $\a$ to the left of $\d$, picking up terms
proportional to $(q-1)$.
Then replace all terms $\a\d$ by $(1+q\b\g)$. Putting
$\g$ to the right of $\b$, $p_1$ finally has the form
either $\a^n\b^x\g^y+(q-1)(...)$ ,
$\d^n\b^x\g^y+(q-1)(...)$ or $\b^x\g^y+(q-1)(...)$ with
$x,y \in \Z, n\in \N$.

Similarly, define a {\em normal form for $p_0$} as
follows: bring all $\b, \g$ to the right of all $\a,
\d$,
order $\b$ to the left of $\g$ and replace all terms
$\b\g$ by $(\a\d-1)/q$. Now order $\a$ to the left of
$\d$
picking up terms prop. to $(q-1)$. $p_0$ finally has the
form
either $\a^x\d^y\b^n+(q-1)(...)$,
$\a^x\d^y\g^n+(q-1)(...)$ or $\a^x\d^y+(q-1)(...)$ with
$x,y \in \Z, n\in \N$.
Now consider the equation
\be
p_0=p_1.
\eq
and put terms in $p_1$ which do not contain inverses to
the left side, in normal
form for $p_0$ (only for monomials which are not
proportional to $(q-1)$, say).
Then let $q=1$ and consider both sides as classical
functions on $SU(2)$.
All terms proportional to $(q-1)$
vanish, and all remaining monomials are in normal form
on both sides and are
easily seen to be independent as functions on $SU(2)$.
This implies that all coefficients are actually zero,
i.e. all terms on both sides are proportional to
$(q-1)$.
(or simply: classical functions
defined on both patches are defined globally on
$SU(2)$). We can now cancel the
greatest common power of $(q-1)$, put regular terms to
the left and apply
the same argument. This cannot go on forever since the
right side can be ordered completely, so both sides must
be zero eventually, proving that $p_0=p_1\in
SU_q(2)$.
Using (\ref{rep_P0}), this immediately shows that
\be
\PP =<{p\in SU_{q}(2): deg(p)=2nk,k\in \Z}>,
\eq
as claimed.
\endproof

The essence of the proof is to write things in the form
("class")+$(q-1)$ ("quantum") and to apply classical
reasoning to ("class"), which should be a fairly general
strategy.
Proposition \ref{intersect} and (\ref{rep_P0})
generalize the result of \cite{BM}
for $n=1$.

We can now introduce the same induced 3-D calculus on
the bundles as in
section 2, i.e. the calculus on the patches $P_i^{(2n)}$
is defined by
\be
N^{(2n)}_{P_i}=P_{i}^{(2n)} N^{(2n)}_P P_{i}^{(2n)},
\eq
with the same ideals as in section 2. Using
$\ \widetilde{\ }=(id\tens\pi)\kappa^{-1}$ in a Hopf
algebra one can easily see
$\widetilde{\ }(N^{(1)}_P)=P^{(1)} \tens M^{(1)}_A$, and
$\widetilde{\ }(N^{(2n)}_P)=\P \tens M^{(2n)}_A$
with a similar argument as in section 2. So
$\P_i$ are trivial QPB with this calculus by example
(4.11) in \cite{BM}.

It was already shown in section 2 that
\begin{eqnarray}
\omega(Z^{kn})=S((\a^{2kn})_{(1)})d(\a^{2kn})_{(2)}
=\kappa(1\otimes (\a^{2kn}-1)), \\
\omega(Z^{-kn})=S((\d^{2kn})_{(1)})d(\d^{2kn})_{(2)}
=\kappa(1\otimes (\d^{2kn}-1))
\end{eqnarray}
for $k\in\N$ defines a connection one - form.
Any monomials of degree $2nk$ in $P_i^{(2n)}$ can be
written in the form
$\Phi_i(Z^{2nk}) B$ or $B \Phi_i(Z^{2nk})$, and so
one can put $\omega$ in the standard form of a
connection
one-form in $\PP_{0}$ and $\PP_{1}$:
\begin{equation}
\omega(a)=\Phi_{i}^{-1}(a)\b_{i}(a)\Phi_{i}(a)+\Phi_{i}^
{-1}(a)d\Phi_{i}(a) ;i=0,1 ;a\in A^{(2n)},
\end{equation}
where $\b_{i}\in \Gamma^{(2n)}_{ihor}$ and
$\b_{i}(1)=0$.

Now let us show the following:

\begin{prop}
The calculus on $\PP,\PP_0,\PP_1$ is admissible
(defined by (\ref{ad1}),(\ref{ad2})).
\end{prop}

\proof
The reasoning is as in the previous proposition.
Assume we have $\omega_0, \omega_1$ in
$\Gamma^{(2n)}_{P_0}$ resp.
$\Gamma^{(2n)}_{P_1}$ with $\omega_0=\omega_1$ in
$\Gamma^{(2n)}_{P_{01}}$.
Since in the 3D - calculus all one - forms on $SU_q(2)$
and thus on $\PP_i$ can
be written in terms of three
left - invariant Maurer - Cartan forms
$\omega^0, \omega^1, \omega^2$ which have simple
commutation relations
\begin{eqnarray}
\omega^0\alpha & = & q^{-1}\alpha\omega^0 , \quad
\omega^0\beta =
q\beta\omega^0, \nonumber \\
\omega^1\alpha & = & q^{-2}\alpha\omega^1 , \quad
\omega^1\beta =
q^2\beta\omega^1, \label{commut.3d} \nonumber\\
\omega^2\alpha & = & q^{-1}\alpha\omega^2 , \quad
\omega^2\beta =
q\beta\omega^2.
\end{eqnarray}
and similarly with the inverses $\a^{-1}$ etc., we can
commute the forms to the
right and have
$\omega_0= f_k \omega^k, \quad \omega_1= g_k \omega^k$
(summation implied), so
\be
f_k \omega^k = g_k \omega^k.
\eq
As in proposition \ref{intersect} put both $f_k$ and
$g_k$ in their respective
normal form ("class") + $(q-1)$("quant") and bring all
regular terms of
$g_k$ to the left side. Then putting $q=1$, the
"classical" parts are all
independent as one - forms since the $\omega_i$ are and
therefore vanish.
Cancelling $(q-1)$ and repeating the argument, it
follows that $\omega_0$ and
$\omega_1$ are elements of $\Gamma^{(1)}_P$ and in fact
in $\Gamma_P^{(2n)}$,
since the degree is conserved.
\endproof

Now we can use theorem \ref{gen.patch}: suppose $\rho
\in
\Gamma^{(2n)}_{0 hor} \cap \Gamma^{(2n)}_{1 hor}$,
so $\rho \in \Gamma^{(2n)}_P$.
We can expand it as above
\be
\rho=f_0\omega^0+f_1 \omega^1+f_2\omega^2,
\eq
with $f_i \in \PP$. But $\omega^0$ and $\omega^2$ are
horizontal (explicitely:
$\omega^0 = \d^2 db_- +q^{-2} \b^2 db_+
-q^{-1}(1+q^{-2})\b\d db_3$ and
$\omega^2 = -\g^2 db_- - q^{-2} \a^2 db_+ +
q^{-1}(1+q^{-2})\a\g db_3$ ),
while $\omega^1$ is not. Therefore $f_1 = 0$, and $\rho
\in \Gamma^{(2n)}_{hor}$,
since all coefficients of $dB$ must have degree $2n$.
So $\PP$ is a QPB with a general differential calculus,
with the same
connection form $\omega$ restricted to elements $a\in
A^{(2n)}$.

Finally we would like to mention that since the
trivializations are not "real"
for $q\neq 1$, one might just go ahead and use
trivializations such as
$\Phi(Z^{1/2}) = \a$ which do not respect the star -
structure even for $q=1$,
at least as computational tools. Since we know that the
"global" bundle with the star - structure does have the
correct classical limit, this may
be an acceptable and useful strategy, and deserves
further consideration.

\section{Concluding Remarks}
\subsection {A note on gauge transformations}
A gauge transformation is a convolution invertible map
$\gamma:A\rightarrow B$:
\begin{equation}
\g \ast \g^{-1} = \g^{-1} \ast \g = 1.
\end{equation}

Let us define the "primitive charge" of a monomial in
$B$ as $(n_{-}-n_{+})$,
where $n_{\pm}$ are the total powers of $b_{\pm}$
appearing in the monomial or equivalently (power of $\a
-$
power of $\d$). This is preserved by the commutation
relations, as our previous degree.
Suppose that $\gamma=\sum_{k=i}^{j}\gamma^{(k)}$ ,where
each
$\gamma^{(k)}$ contains only monomials that have
primitive charge $k$.
Hence $i$ and $j$ are the minimum and maximum of the
primitive charges of all monomials in $\gamma$.
Let the convolution inverse of $\gamma$ be denoted in
the same way:
$\gamma^{-1}=\sum_{k=i\prime}^{j\prime}\gamma^{(k)}\prime$.
So
\begin{equation}
1=\epsilon(Z^{n/2})\cdot 1=\gamma \ast
\gamma^{-1}=\sum_{k=i+i\prime}^{j+j\prime}
\gamma^{(k)}\prime\prime \label{gamma}
\end{equation}
which implies that $i+i\prime=j+j\prime=0$.
The only possibility that this can be true is that
$i=j=-i\prime=-j\prime$, which means that all monomials
in $\gamma$ have the same
primitive charge $n$. However, this means in the
classical limit that $\g$ is proportional to
$e^{in\phi}$.
That is, by admitting only finite sums in a convolution
invertible $\g$ one is restricting oneself to a very
special, rigid
class of gauge transformations. Thus infinite series
cannot be avoided in general.

\subsection{Remarks on the Chern Class}
Classically, the monopole charge $n$ is given by an
integration over the base
of the first Chern class
$$\frac{1}{2 \pi i} \int_{S^2} F =n,$$
where $F=dA_+=dA_-.$ Here $A_+, A_-$ are the connection
form on the northern
and southern hemisphere respectively and the global
connection form is given in
terms of trivalizations as
\be
\om= \cases {A_+ + i d\varphi_+ , on H_+&\cr
             A_- + i d\varphi_- , on H_-}
\eq
with $e^{i \varphi_{+,-}}$ being the local
trivalization.

In the deformed case, we have the global connection form
$\om$. Suppose it is written in terms of
trivialization as \cite{BM}
$$ \om = \phi^{-1}_i \b_i \phi_i +\phi_i^{-1} d\phi_i,
$$
then it is not hard to check that
\be
d\om = \phi^{-1}_i(d\b_i +\b_i \b_i ) \phi_i = d\omega +
\omega \omega,
\eq
which is in fact the curvature 2-form on $P$
(\cite{Hajac}, cp. \cite{BM}).
Carrying the $\phi_i$ through the $d\om$, we get
\be d\b_i +\b_i \b_i  = q^{2n} d\om =q^{2n}[n]_{q^{-2}}
d\om^1 \eq
which is again equal for the two patches and explicitely
horizontal. It leads us
to define the deformed Chern class as
\be
\frac{1}{2 \pi i} F=\frac{1}{2 \pi i} (d\b_i +\b_i \b_i)
=\frac{q^{2n}}{2 \pi i} [n]_{q^{-2}} d\om^1.
\eq

Consider the base $B=S^2_q=<b_+, b_-, b_3> \subset
SU_q(2)$, with the calculus
inherited from the 3-D calculus on $ SU_q(2)$. Denote
$\Gamma_B =BdB$ and
introduce the set $\Gamma ^{\wedge^2}_{B}$ of 2 forms on
$B$. Notice that
$\Gamma ^{\wedge^2}_{B}$  contains elements of the form
$Bdb_i db_j, i,j=-,+,3.$
Since
$$ db_+=\g^2 \om^0 -q^2 \d^2 \om^2,$$
$$ db_-=\a^2 \om^0 -q^2 \b^2 \om^2,$$
\be db_3=\a\g \om^0 -q^2 \b\d \om^2.\eq
So, $\Gamma ^{\wedge^2}_{B}=B \om^0 \om^2= Bd\om^1$.
Because $d\om^1$ is a central element in $B_{01}$,
under a gauge transformation $U \in B_{01}$ we have
$F \rightarrow U^{-1}FU = F$.

Notice that $\om^0 \om^2$ is manifestly left invariant
under the coaction
of $SU_q(2)$, and is the unique top 2-form on $B$.
This allow us to introduce a linear functional
$$ \int_B: \Gamma ^{\wedge^2}_{B} \rightarrow \compl, $$
\be \int_{S^2_q} a d\om^1 = 2 \pi i <a> _{SU_q(2)},
\forall a \in B, \eq
where  $<  > _{SU_q(2)}$ is the invariant "Haar" measure
on $SU_q(2)$\cite{Z}.
This integral is obviously left- and right- invariant
under the coaction of
$SO_q(3)$ and unique as such.
The normalization is choosen to give the correct
classical limit. Classically,
$d\om^1 =i/2 d \Omega.$

Therefore the deformed monople charge is obtained as in
the classical case
\be
\frac{1}{2 \pi i} \int_{S_q^2} F
= \frac{q^{2n}}{2 \pi i} \int_{S_q^2} [n]_{q^{-2}}
d\om^1=q^{2n} [n]_{q^{-2}}.
\eq
This is actually gauge - invariant in the sense that it
does not depend on the
trivialization chosen, but this appears to be the case
only for our particular
connection.

\section{Acknowledgements}
We wish to express our gratitude to Prof. Bruno Zumino
for many useful
discussions, encouragement and support, in particular
for drawing our attention
to this problem. We also thank Arne Schirrmacher for
suggestions, and T. Brzezinski
for sending us \cite{BM_err}.
This work was supported in part by the Director, Office
of
Energy Research, Office of High Energy and Nuclear
Physics, Division of
High Energy Physics of the U.S. Department of Energy
under Contract
DE-AC03-76SF00098 and in part by the National Science
Foundation under
grant PHY-90-21139.

\end{document}